# High-quality Tungsten-doped Vanadium Dioxide Thin Films Fabricated in an Extremely Low-oxygen Furnace Environment

Vishwa Krishna Rajan, Ken Araki, Robert Y. Wang, Liping Wang[*]

School for Engineering of Matter, Transport and Energy, Arizona State University, Tempe, Arizona 85287, USA

*Corresponding Author. Email: Liping.Wang@asu.edu*

## Abstract

This work reports the fabrication and characterization of high-quality tungsten-doped vanadium dioxide ($W_xV_{1-x}O_2$, $x = 0$~$3$ at. %) by thermal oxidation of sputtered tungsten-vanadium alloyed thin films with different atomic percentages and high-temperature annealing in an extremely low oxygen atmosphere (5 to 20 ppm) along with reduction of surface over-oxides in high vacuum (1 mPa). Oxidation parameters such as temperature, time and nitrogen purging rate are first optimized for obtaining high quality undoped $VO_2$ thin film. Insulator-to-metal (IMT) phase transition behavior of $VO_2$ thin films fabricated in a low-$O_2$ environment is characterized with temperature dependent spectral infrared transmittance and electrical resistivity measurements, where there is 15% higher infrared transmittance change and additional 1 order change in resistivity in comparison with $VO_2$ thin films fabricated in a $O_2$-rich environment. Grazing angle X-ray diffraction scan confirms no presence of higher oxides in the $VO_2$ oxidized in low-$O_2$ environment, which improves its quality significantly. Comprehensive studies on thermal annealing and vacuum reduction for tungsten doped $VO_2$ thin films are also carried out to find the optimal fabrication conditions. With the tungsten at. % measured by X-ray photoelectron spectroscopy, the optimal $WVO_2$ thin films fabricated through this streamlined oxidation, annealing and reduction processes in extremely low-$O_2$ furnace environment exhibit lowered IMT temperature at −23°C per at.% of tungsten dopants from 68°C without doping. This low-cost and scalable fabrication method could facilitate the wide development of tunable $WVO_2$ coatings in thermal and energy applications.

# 1. Introduction

Thermochromic vanadium dioxide ($VO_2$) has attracted significant attention due to its reversible phase transition from insulating (monoclinic) phase to metallic (tetragonal) phase (IMT) at 68°C [1]. This unique behavior that leads to dramatic changes in its optical and infrared properties has found important applications in smart windows[1,2], adaptive radiative cooling [3–5], thermal rectification [6–8], and thermal camouflage [9,10]. Scalable and cost-effective fabrication of high-quality $VO_2$ thin films poses a significant challenge since vanadium oxides can exist in multiple oxidation states. Presence of other oxides within the film could curtail the optical properties and phase transition behaviors of any $VO_2$ based device in the intended temperature range. Common techniques to fabricate $VO_2$ thin films include chemical vapor deposition [11,12], reactive magnetron sputtering [13–15], pulsed laser deposition (PLD) [6,16–18], atomic layer deposition (ALD) [19–21], sol-gel method [22–24] and molecular beam epitaxy (MBE) [25,26]. High-quality $VO_2$ thin films are usually obtained in the above-mentioned techniques by precisely controlling oxygen partial pressure in a vacuum atmosphere at elevated temperatures. In addition, CVD and ALD require carefully chosen precursors that could complicate the fabrication procedures. MBE, which could yield best quality epitaxial $VO_2$ thin films, requires lattice-matching substrate such as sapphire and is expensive to operate.

Few works reported growth of $VO_2$ thin films via thermal oxidation [27–29] in a $O_2$-rich atmosphere, where it suffers from excessive surface over-oxides deteriorating the phase transition performance. On the other hand, vanadium oxidation state could reduce in an extremely low-$O_2$ environment at high temperatures due to creation of oxygen vacancy defects in vanadium oxide crystal lattice. Several works have demonstrated the reduction of vanadium oxides from higher to lower oxidation or even back to metal vanadium at elevated temperatures in extremely low-$O_2$ atmosphere created by high vacuum [30–34], Ar [35–37] and $N_2$ [38] environment. $VO_2$ thin films were also obtained by reducing $V_2O_5$ via reactive sputtering [13], sol-gel [34] and PLD[18] techniques at similar temperature and low-$O_2$ atmosphere conditions.

As much as it is important to control the stoichiometry of $VO_2$ thin films, it is also crucial to lower its phase transition temperature for room-temperature or space applications. Doping with metal ions such as Mg [39,40], Nb [41–43], Mo [42,43] and W [44,45] is widely used since it distorts V-V dimer in the monoclinic phase thereby altering the crystal structure. For instance, introducing W atoms into $VO_2$ crystal lattice could lower the phase transition temperature by 15 to 25°C per at.% [44–47]. Applications such as radiative cooling [48] and spacecraft thermal control [6,49] require phase transition temperature near room temperature where >2 at.% of tungsten doping is necessary. Diffusion of tungsten atoms within the $VO_2$ lattice is crucial for lowering transition temperature, which could be achieved by high-temperature annealing [50–52]. While tungsten-doped $VO_2$ films ($WVO_2$) could be fabricated via ALD [46], PLD [53] and sol-gel methods [54], there is an urgent need for a more cost-effective and scalable approach for fabricating high-quality tungsten-doped $VO_2$ thin films.

This work aims to demonstrate the low-cost fabrication and characterization of high-quality $WVO_2$ thin films grown in extremely low-$O_2$ furnace environment from subsequent sputtering, oxidation, annealing, and vacuum reduction processes. Tungsten-vanadium alloyed thin films of different doping levels are sputtered on the 2-inch undoped silicon (UDSi) and quartz wafers. Effects of oxidation temperature, time and nitrogen gas purging rate are studied to find the optimal oxidation conditions for undoped $VO_2$ thin films. Temperature-dependent infrared transmittance and electrical resistivity measurements are performed to characterize the phase transition behaviors in comparison to the $VO_2$ film grown in $O_2$-rich environment. X-ray diffraction (XRD) is conducted to study the vanadium oxide states grown in both low-$O_2$ and $O_2$-rich furnace conditions. Comprehensive studies on the annealing temperature and vacuum reduction time are also carried out for $WVO_2$ of different doping levels. Finally, the IMT behaviors of optimally grown $WVO_2$ thin films are characterized with doping levels measured by X-ray photoelectron spectroscopy (XPS).

## 2. Methods

Figure 1 depicts the fabrication flowchart of undoped and tungsten-doped $VO_2$ thin films in extremely low-$O_2$ furnace environment as well as materials characterizations for film compositions and IMT behaviors. About 25-nm-thick vanadium or tungsten-vanadium thin films are first sputtered onto 280-μm-thick double-side polished infrared-transparent UDSi wafers in 2-inch diameter (resistivity $\rho$>10,000 Ω·cm, orientation <100>, University Wafers Inc.). Vanadium thin film is then oxidized into high-quality $VO_2$ thin film in an extremely low-oxygen furnace environment by purging with $N_2$ gas at a proper flowrate, while tungsten-vanadium films are additionally annealed at higher temperatures after low-$O_2$ oxidation followed by surface reduction of over-oxides in high vacuum to obtain high-quality tungsten-doped $VO_2$ thin films. XRD scans are conducted to study the oxide states and crystallinity of undoped $VO_2$ films, while XPS characterization is carried out to determine the tungsten doping levels within the tungsten-doped $VO_2$ thin films. Temperature-dependent infrared transmittance and electrical resistivity measurements are performed to reveal the IMT behaviors of undoped and tungsten-doped $VO_2$ thin films optimally grown in the low-$O_2$ furnace environment.

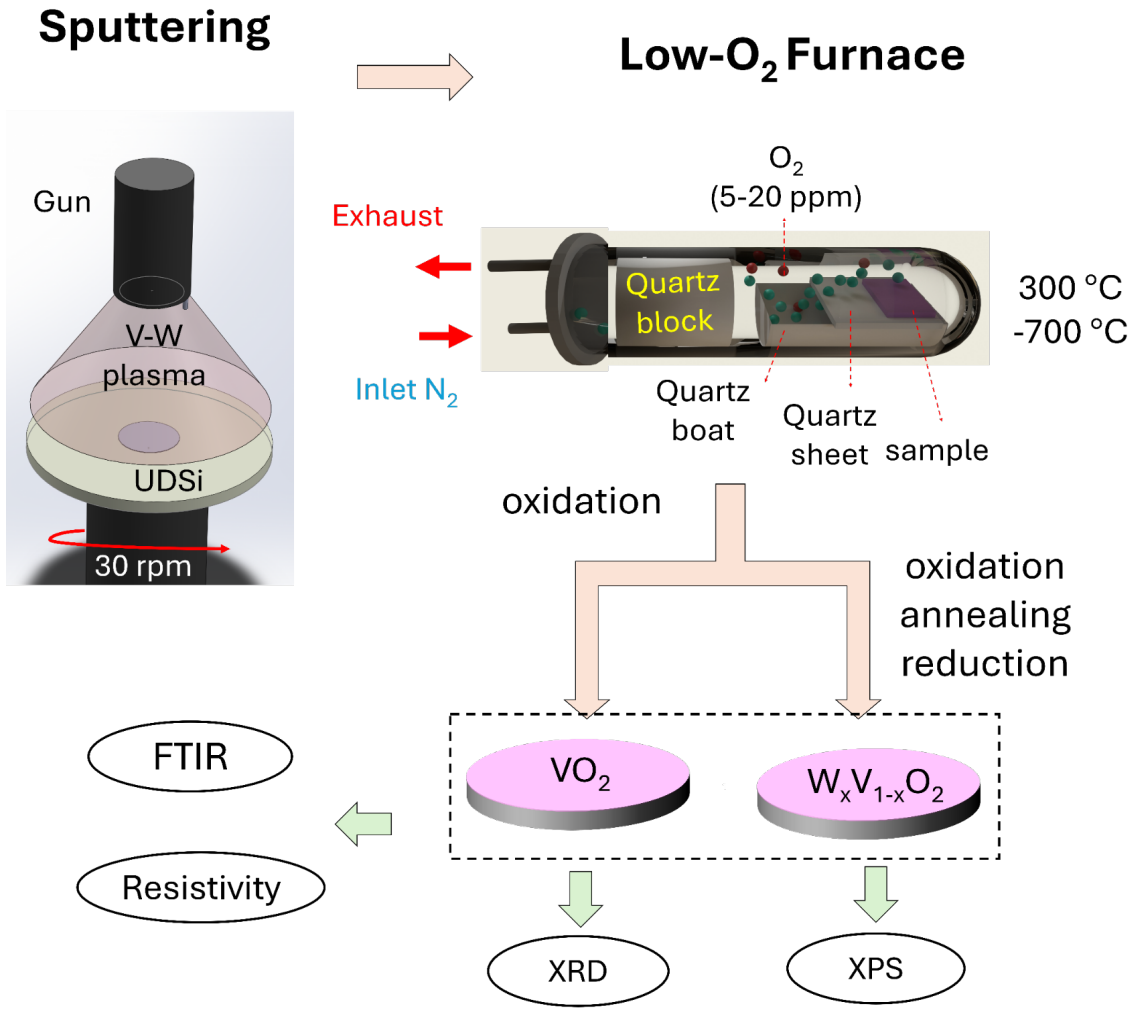

**Figure 1.** Flowchart to fabricate and characterize high-quality undoped and tungsten-doped $VO_2$ thin films thermally grown in extremely low-$O_2$ environment.

2.1 Sputtering of metal films

Vanadium or tungsten-vanadium thin films are sputtered onto 2-inch UDSi substrates from pure vanadium (99.99 % purity, Kurt J. Lesker) or tungsten-vanadium alloy sputter targets with ~1, ~2, ~3 at.% of tungsten (±0.5 at.%, 99.9% purity, Stanford Advanced Materials).Once the sputtering chamber is pumped down to the base pressure of $5\times10^{-6}$ Torr, argon gas is introduced to maintain the chamber pressure constantly at 4 mTorr throughout the entire deposition. Magnetron gun is then turned on by a DC power supply at about 120 W to sputter the target, while the deposition rate is measured by the crystal monitor after calibration. To preclean the target and achieve a stable deposition rate, the plasma is kept on for 10 minutes with the substrate shutter closed. Once a stable deposition rate of 1.0 Å/s is reached, substrate shutter is opened to allow the deposition onto the UDSi wafers. The platen is rotated at 30 rpm throughout the entire process for the uniform deposition of the thin films. The substrate shutter is closed once the thickness from the crystal monitor reads 250 Å. The gun power is then turned off and the sample is allowed to cool down for at least 10 minutes before the chamber is vented to atmosphere.

2.2 Low-$O_2$ furnace setup

A hybrid furnace made of a single-ended quartz tube of 50-mm diameter inserted into a 4-inch cubic muffle furnace (KSL1100XSH, MTI Corporation) with $N_2$ gas purging and vacuum pumping is used to create the extremely low-$O_2$ environment (5~20 ppm) for oxidation, annealing and surface reduction to grow high-quality undoped and tungsten-doped $VO_2$ thin films. The furnace temperature ranges from 200°C to 1000°C with accuracy of ±1°C with precise proportional-integral-derivative (PID) control. The quartz tube is equipped with a flange where $N_2$ gas (99.99 % purity) is introduced at a few psia with its flowrate controlled by a rotameter (SKUW-493607, Aalborg Instruments) between 0.5 to 4.8 liter per minute (lpm) with 0.1 lpm resolution. The exhaust gas exiting the quartz tube passes through a trace oxygen sensor (TO2-1x, Southland Sensing Ltd) which could detect $O_2$ concentration from 0.01 ppm up to 25% recorded by an oxygen analyzer every 4 seconds during the process. To facilitate the surface reduction, a turbomolecular vacuum pump (Hicube 80, Pfeiffer Vacuum) is also connected to the exhaust, which could bring down the quartz tube to 1 mPa with inlet closed. Sputtered metal thin film samples are placed on a quartz sheet on top of a quartz boat sitting around one inch away from the end of the quartz tube where uniform temperature exists. A complete furnace process for growing tungsten-doped $VO_2$ thin films includes five stages of heating, oxidation, annealing, reduction, and cooling at different furnace temperatures and duration, which are programmed via the furnace temperature controller for a streamlined process. The photo of the furnace and schematic describing the gas lines is shown in section S1 of Supplemental Information.

2.3 Temperature-dependent infrared spectral measurements

Spectral infrared transmittance of undoped and tungsten-doped VO$_2$ thin films on UDSi wafers is measured between 2 to 20 μm in a wide temperature range between −30°C and 100°C with a Fourier-transform infrared (FTIR) spectrometer (Nicolet iS50, Thermo Fisher Scientific) at normal incidence. Samples are placed on a home-built temperature stage made of a copper sample holder with an 8-mm aperture, a Peltier element (TR060-6.5-40-03LS, Coherent Thermal Solutions), and a water block heat sink. A K-type thermocouple is attached to the copper sample holder next to the sample with thermal paste, and a PID temperature controller (CSi8D, OMEGA Engineering) modulates the power supplied to the Peltier element such that the sample temperature is maintained at the setpoint with ±1°C temperature stability. For temperatures less than 5°C, the FTIR sample compartment is purged with dry N$_2$ gas to prevent the formation of water and ice layers on the sample surface. Spectral measurement with 32 scans at 4 cm$^{-1}$ resolution is taken once the sample temperature reaches the setpoint for 5 mins to ensure the steady state is achieved. Temperature-dependent spectral reflectance measurements are done similarly with a custom-built temperature stage and a 10-deg specular reflection accessory (10Spec, PIKE Technology). Please see more details on the setup and validation of FTIR measurements in the Section S2 of Supplemental Information.

2.4 Temperature-dependent four-probe resistivity measurement

The temperature-dependent resistivity of the undoped and tungsten-doped VO$_2$ thin films on UDSi wafers is measured using a home-built four-probe setup with a similar temperature stage. Voltage-current curves are obtained at a given temperature with a source meter (2401, Keithley), and the resistivity can be obtained by $\rho(T) = C\pi S(T)t/\ln2$ where $C$ is the geometrical parameter, $S$ is the slope of V-I curve and $t$ is thickness of the sample. Resistivity is then normalized to the lowest value in the metallic phase as $\rho(T)/\rho_m$ to show the temperature effect only on the IMT behavior of WVO$_2$ thin films. Please see more details on the setup and validation measurements with undoped and heavily doped silicon wafers in the Section S3 of Supplemental Information.

2.5 XRD and XPS characterizations

Grazing-incidence X-ray diffraction is respectively performed at the surface and within the film by fixing the angle of incidence as 0.5° and 2° with Cu K$\alpha$ X-ray source. The diffraction pattern is matched with powder diffraction databases generated with HighScore Plus software to obtain the vanadium oxide states and assess the crystallinity of optimally grown undoped $VO_2$ thin films. X-ray photoelectron spectroscopy (XPS) surface scans are conducted with Kratos Axis Supra+ with Al k$\alpha$ 1486.6 eV X-ray source. V2p elemental scans from binding energies 510.0 eV to 540.0 eV and W4f elemental scans from binding energies 30.0 eV to 50.0 eV are performed to measure the atomic concentration of W for the optimally grown $W_xV_{1-x}O_2$ thin films. Raw data is analyzed and fitted with CasaXPS software with Shirley algorithm utilized to draw background for both V2p and W4f elemental scans. Lorentzian asymmetric curve is used to fit the V2p, V3p and O1s peaks while Gaussian Lorentzian peaks are used to fit the peaks of W4f. XPS scans are done at the center of the films after the films are etched with 5 keV $Ar^+$ ions for 5 minutes.

## 3. Results and Discussion

3.1 Improved $VO_2$ thin film quality via low-$O_2$ oxidation

To obtain the best quality of thermally grown undoped $VO_2$ thin films in low-$O_2$ environment, a series of furnace oxidation tests is carried out to find optimal oxidation temperature, duration, and $N_2$ gas flowrate. While the quality of the $VO_2$ thin films could be slightly affected by the heating and cooling conditions, the oxidation parameters play a much more important role (see Section S4 in the Supplemental Information for parametric study during heating stage). Therefore, the heating ramp rate is fixed at 10°C/min, while the $N_2$ gas at 4.5 lpm is purged through the quartz tube to minimize oxidation during both heating and cooling stages.

Effect of oxidation temperature is studied first with the oxidation time fixed at 3 hours and $N_2$ gas flowrate of 1.0 lpm during oxidation. Figure 2(a) shows the infrared spectral transmittance measured at 25°C for the insulating phase and at 90°C for the metallic phase after oxidizing 25-nm vanadium thin films on UDSi substrate at different oxidation temperatures from 300°C to 700°C. The infrared transmittance is around 25% with little changes between 25°C and 90°C measurements after 300°C oxidation, suggesting that the temperature is not high enough to oxidize most of the vanadium film. With the oxidation temperature increases to 400°C, typical $VO_2$ phase transition is observed after the oxidation with the transmittance in the insulating phase around 53% and that in the metallic phase around 20%, confirming that vanadium film is fully oxidized. Note that the major transmittance dip around 16.6 μm wavelength is associated with silicon phonon absorption from the UDSi substrate. Noticeably, there is a couple of minor peaks around 10 μm wavelength possibly due to stretching modes of vanadium oxides also observed by others [13,55][56]. With 500°C oxidation, similar typical VO$_2$ phase transition is also observed with slightly enhanced near-infrared transparency in the insulating phase and slightly lowered infrared transmittance in the metallic phase by 2%, indicating optimal oxidation temperature of 500°C to obtain largest transmittance change upon VO$_2$ IMT. When the oxidation temperature is further increased to 600°C, the film exhibits much higher transmittance reaching 40% in its metallic phase as most of the film is over-oxidized into higher vanadium oxides such as V$_2$O$_5$. Finally, with 700°C oxidation the film loses the phase transition behavior completely with almost the same high transmittance of 53% measured at both 25°C and 90°C as it is over-oxidized entirely. Note that all the oxidation tests are done in extremely low-O$_2$ environment as shown by the O$_2$ ppm level in Figure 2(b). The O$_2$ concentration drops down to 5 ppm at the end of heating stage with 4.5 lpm N$_2$ purging, then it stays around 12~18 ppm with 1.0 lpm N$_2$ purging during the entire oxidation stage, and finally it drops to 5 ppm again with 4.5 lpm N$_2$ purging during cooling.

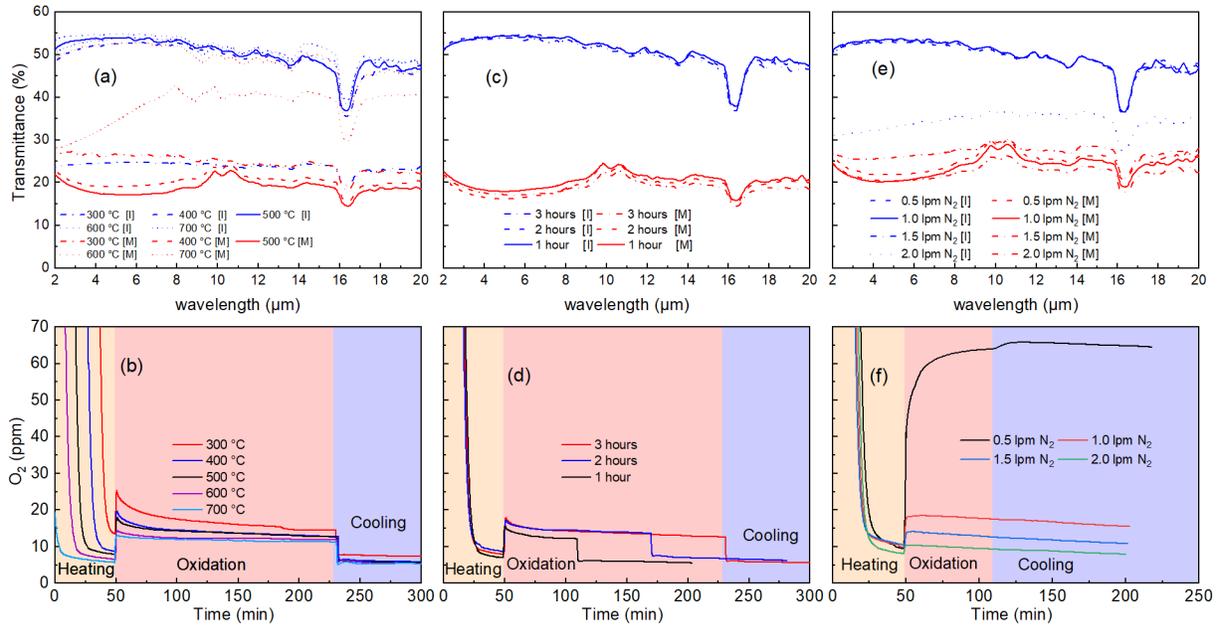

**Figure 2.** Low-O$_2$ oxidation tests for thermally growing 50-nm undoped VO$_2$ thin films: (a, b) effect of oxidation temperature, (c, d) effect of oxidation time, and (e, f) effect of N$_2$ gas flowrate, with (a, c, e) infrared spectral transmittance measured at 25°C for its insulating phase [I] and at 90°C for its metallic phase [M] along with (b, d, f) O$_2$ ppm level measured during each test.

With 500°C temperature and 1.0 lpm N$_2$ purging flowrate during oxidation, several 25-nm vanadium films of about 0.5 inch$^2$ in size are oxidized for different durations of 1, 2 and 3 hours.

As shown in Figure 2(c) and 2(d) all three films are oxidized consistently in ~15 ppm $O_2$ environment with the same typical $VO_2$ IMT behavior in infrared transmittance. This confirms that 1 hour oxidation time is sufficient to achieve optimal oxidation, and more importantly that, excessive oxidation time will not over-oxidize the film in such a low-$O_2$ environment. This is advantageous over previous furnace growth of $VO_2$ thin film in $O_2$-rich environment, where precise control of oxidation time is required to avoid under or over-oxidation of the film. On the other hand, it is also observed that larger samples would need longer oxidation time to get fully oxidized due to extremely low concentration of oxygen in the furnace (see Section S5 in the Supplemental Information for more details). Effect of $N_2$ gas flowrate which is expected to change the $O_2$ ppm level during oxidation is studied at last with 1-hour oxidation time and 500°C oxidation temperature. As shown in Figure 2(e), the $VO_2$ films after oxidization with 0.5, 1.0 and 1.5 lpm $N_2$ gas flowrates exhibit almost the same infrared transmittance spectra with typical IMT behavior, while the $O_2$ concentration varies from 65 to 10 ppm from Figure 2(f), suggesting that high quality of $VO_2$ thin film can be grown within this $O_2$ ppm range. However, with 2 lpm $N_2$ gas purging, the $O_2$ concentration during oxidation drops below 10 ppm, and much lower infrared transmittance at its insulating phase only around 35% is observed, suggesting that the vanadium film is not completely or under oxidized at the sub-10-ppm $O_2$ level. Finally, 500°C, 1 lpm $N_2$ purging rate, and 1 hour are taken as optimal oxidation parameters for oxidizing 25-nm vanadium thin films of 0.5 inch$^2$ in low-$O_2$ furnace atmosphere into high-quality $VO_2$. Note that the fabrication method is scalable as $VO_2$ thin films have been successfully grown on 2-inch UDSi wafer with excellent uniformity as well as on other substrate materials such as quartz (see Section S5 and S6 in the Supplemental Information for more details).

      To illustrate the improved quality of $VO_2$ thin film grown in low-$O_2$ furnace atmosphere, another 25-nm vanadium film is oxidized in $O_2$-rich environment (7 vol.%) at 300°C for 4 hours following our previous furnace process [27]. Temperature-dependent infrared transmittance of both $VO_2$ thin films is measured from 30°C to 95°C at 5°C intervals outside the phase transition region and at 2°C intervals within the phase transition (see Section S7 in the Supplemental Information). Figure 3(a) shows the heating/cooling curves of infrared transmittance at 8 μm wavelength of both 50-nm $VO_2$ thin films. Upon IMT phase change, the $VO_2$ thin film grown in low-$O_2$ condition achieves much larger transmittance change of 37%, while $VO_2$ grown in $O_2$-rich environment could only achieve 8% change in transmittance. The IMT behavior is also further

compared on the transmittance derivative with respect to temperature ($d\tau/dT$) for both VO$_2$ thin films as shown in Figure 3(b) upon heating or cooling. The VO$_2$ thin film grown in low-O$_2$ environment exhibits up to 4% transmittance decrease per 1°C temperature change, while that grown in O$_2$-rich environment could only change transmission by less than 0.5% per 1°C with IMT phase transition shifted slightly to lower temperatures along with wider thermal hysteresis between heating and cooling. Temperature-dependent electrical resistivity is also measured for the IMT behaviors of both VO$_2$ thin films. As shown in Figure 3(c), the resistivity of the VO$_2$ thin film grown in low-O$_2$ condition exhibits 60 folds change upon IMT, while that of VO$_2$ thin film grown in O$_2$-rich environment only changes by 3 times. The resistivity derivatives with respect to temperature $d\log_{10}(\rho/\rho_m)/dT$ upon heating and cooling in Figure 3(d) presents similar behavior as the transmittance derivatives, confirming much improved VO$_2$ thin film quality with much sharper phase transition grown in low-O$_2$ condition.

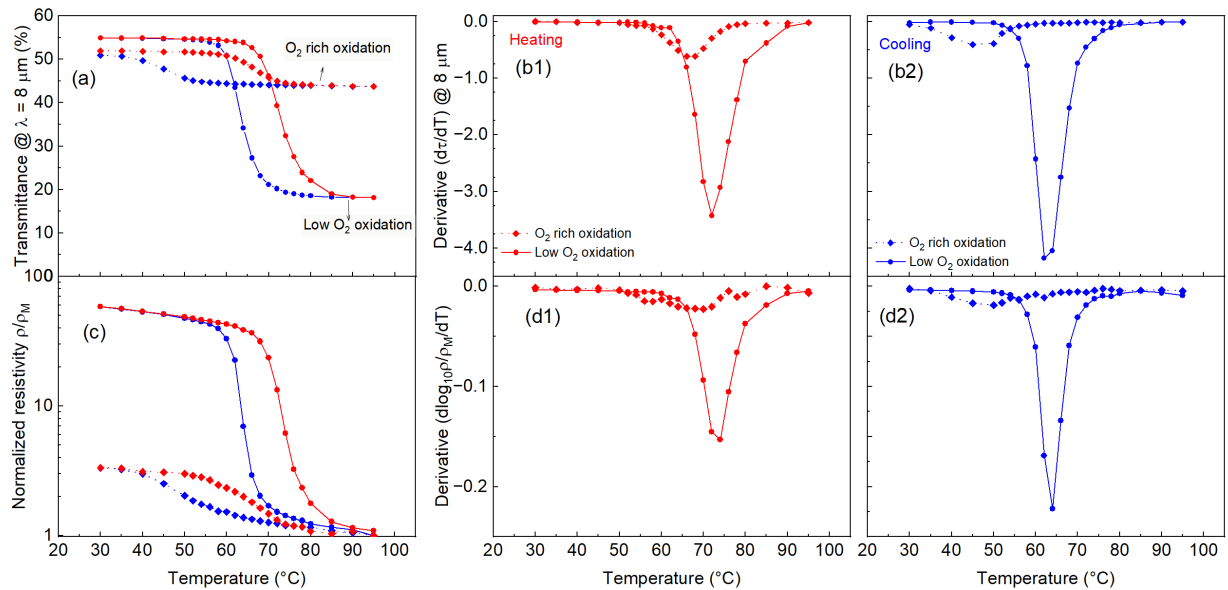

**Figure 3.** Insulator-to-metal transition behaviors of 50-nm VO$_2$ thin films thermally grown in O$_2$-rich and low-O$_2$ furnace environment with (a) temperature-dependent infrared transmittance at 8 mm wavelength and (b) its derivative with respect to temperature upon heating and cooling, as well as (c) temperature-dependent electrical resistivity normalized to the lowest value in metallic phase and (d) its derivative with respect to temperature upon heating and cooling.

It is hypothesized that, the degraded quality of VO$_2$ thin film grown in O$_2$-rich furnace environment is associated with over-oxides such as V$_2$O$_5$ formed at the film surface in the presence of high oxygen concentration, which is precisely controlled in low-O$_2$ environment with N$_2$ gas purging and trace $O_2$ monitoring such that pristine $VO_2$ is obtained with superior IMT properties. This is confirmed by the grazing-incidence XRD scans for vanadium oxide states and phases in both $VO_2$ thin films grown in $O_2$-rich and low-$O_2$ furnace environment. As shown in Figure 4(a) at the film surfaces (incidence angle $\omega = 0.5°$), presence of over-oxides $V_2O_5$ ($2\theta = 20.3°, 21.7°, 37.0°, 50.9°$) and $V_4O_9$ ($2\theta = 24.4°$) without $VO_2$ is observed for $VO_2$ oxidized in $O_2$-rich atmosphere, whereas only $VO_2$ ($2\theta = 28.0°, 40.0°, 55.8°, 57.8°$) without any over-oxides are present for $VO_2$ grown in low-$O_2$ environment. While $VO_2$ and fewer over-oxides are observed within the film (incidence angle $\omega = 2°$) in Figure 4(b), the existence of the over-oxides grown in $O_2$-rich atmosphere is undoubtedly responsible for the deterioration of $VO_2$ infrared properties especially in the metallic phase and its phase transition behavior. As it is clearly demonstrated here, with thermal oxidation in well-controlled low-$O_2$ environment, pristine $VO_2$ is formed at and within the film with exceptional infrared and phase-transition properties.

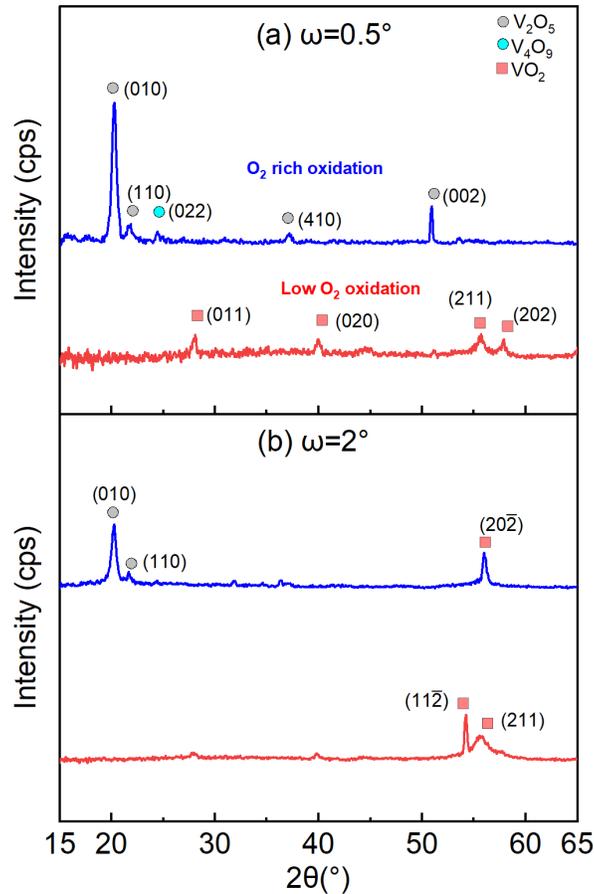

**Figure 4.** Grazing-incidence XRD scans for 50-nm $VO_2$ thin films thermally grown in $O_2$-rich and low-$O_2$ furnace environment: (a) at the surface, and (b) within the film.

3.2 Growth of tungsten-doped VO$_2$ thin films with thermal annealing and surface reduction

Three 25-nm tungsten-vanadium thin films sputtered from alloy targets of different tungsten doping levels are also fully oxidized in the low-O$_2$ furnace environment at 500°C with 1 lpm N$_2$ purging. However, as shown in our previous work [50] and similar works by others [51,52], thermal annealing at a higher temperature is required to fully diffuse the tungsten dopants into VO$_2$ crystal lattice to effectively lower the phase transition temperature. On the other hand, higher temperature would also lead to excessive over-oxidation as shown in Fig. 2(a). While a much higher N$_2$ purging rate could slightly mitigate the over-oxidation during annealing, the cold N$_2$ gas actually decreases the sample temperature which adversely results in insufficient annealing for highly doped WVO$_2$ films. Therefore, a slightly increased N$_2$ purging rate of 1.5 lpm is used during high-temperature annealing, followed by high vacuum process of 1 mPa at 450°C to reduce the surface over-oxides. A comprehensive study is carried out to find the optimal annealing temperature and optimal reduction time for the WVO$_2$ thin films.

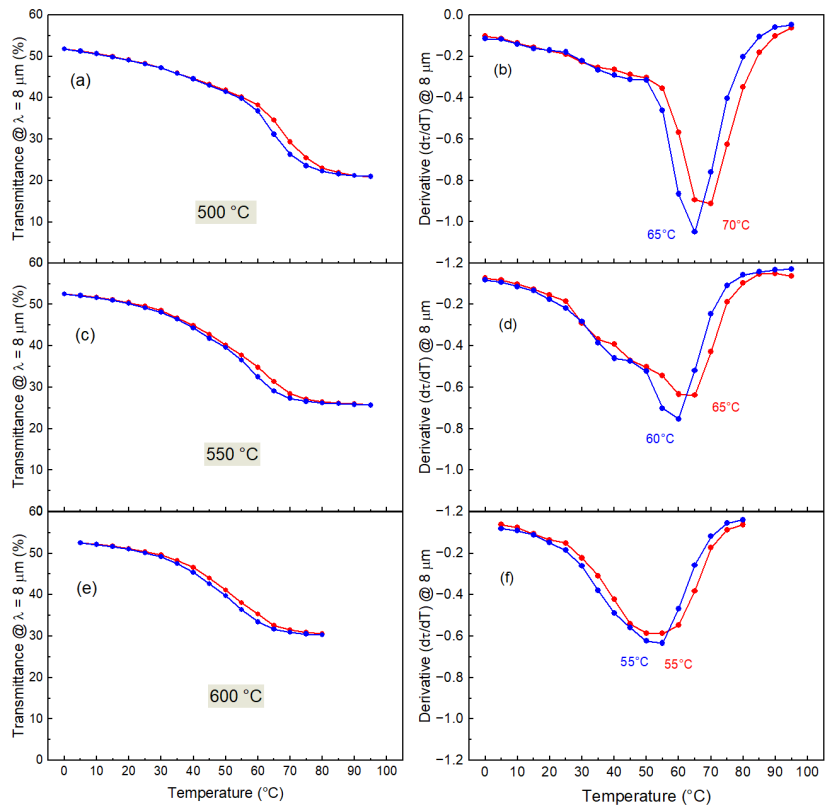

**Figure 5.** Annealing temperature study for ~1 at.% WVO$_2$: (a, c, e) heating and cooling curves from infrared transmittance at 8 μm wavelength and (b, d, f) corresponding derivatives with respect to temperature for samples annealed at (a, b) 500°C, (c, d) 550°C, and (e, f) 600°C.

Figure 5 shows the annealing temperature effect for ~1 at.% WVO$_2$ thin film in terms of infrared transmittance at 8 μm wavelength upon heating and cooling as well as its corresponding derivative. Increasing the annealing temperature from 500°C to 550°C, the IMT midpoint (i.e., temperature at which maximum derivative exists) shifts from 70°C to 65°C during heating and from 65°C to 60°C during cooling. Further increasing the annealing temperature to 600°C, the IMT midpoint is further lowered to 55°C during both heating and cooling with shortened phase transition. However, after 600°C annealing, the infrared transmittance in the metallic phase measured at 95°C is increased with a peak value approaching 35% around 10μm wavelength, suggesting excessive over-oxidation during the high-temperature annealing. To compensate that, the furnace is kept at 450°C and pumped down to high vacuum of 1 mPa for reducing the surface over-oxides. As shown in Fig. 6(a), vacuum reduction of 1 hour could only achieve slight improvement in lowering the short-wavelength transmittance, while reduction of 2 hours and 3 hours could bring down the entire infrared transmittance in the metallic phase effectively by nearly 10%, indicating successful reduction of surface over-oxides. This can be more clearly observed in the transmittance derivative curves in Fig. 6(b) after the ~1 at.% WVO$_2$ reduced by different amount of time in vacuum. With sufficient vacuum reduction of 2 hour, the derivative at the IMT midpoint could reach –0.85 K$^{-1}$ compared to –0.62 K$^{-1}$ without reduction. Reduction for longer time of 3 hours yields about the same results with 2-hour reduction.

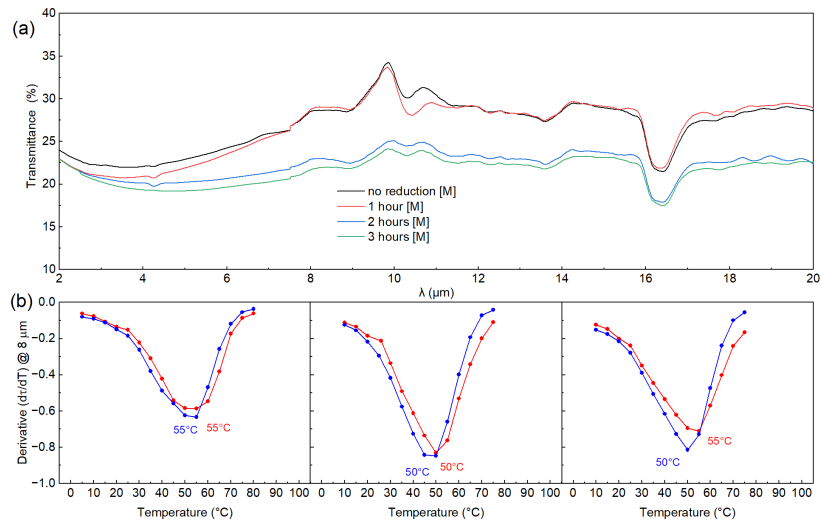

**Figure 6.** Reduction time study at 450°C in high vacuum (1 mPa) for ~1 at.% WVO$_2$ after 600°C and 2 hours annealing: (a) infrared transmittance of metallic phase (measured at 95°C) and (b) transmittance derivative with respect to temperature at 8 μm wavelength upon heating and cooling without reduction and after reduction of 2 and 3 hours.

For heavily-doped WVO$_2$ with doping level higher than > 2 at.%, our previous study found that it requires higher temperature at least 650 °C for fully annealing [50]. Therefore, ~2 at.% WVO$_2$ sample is annealed at 650 °C for 2 hours followed by the vacuum reduction at 450 °C for 3 hours. As shown in Fig. 7(a), the transmittance derivative at 8 μm wavelength shows a clear dip at 35 °C during heating and 30 °C during cooling, indicating that the sample is fully annealed. However, the ~3 at.% WVO$_2$ sample exhibits double dips around 10 °C and 55 °C in the transmittance derivative after 650 °C annealing, suggesting insufficient annealing, as shown in Fig. 7(b). After increasing the annealing temperature further to 675 °C, a single transmission derivative dip is observed which confirms fully annealing of the ~3 at.% WVO$_2$ sample. Note that studies on annealing flowrate and reduction time for ~2 and ~3 at.% WVO$_2$ sample can be found in Section S9 and S10 of Supplemental Information.

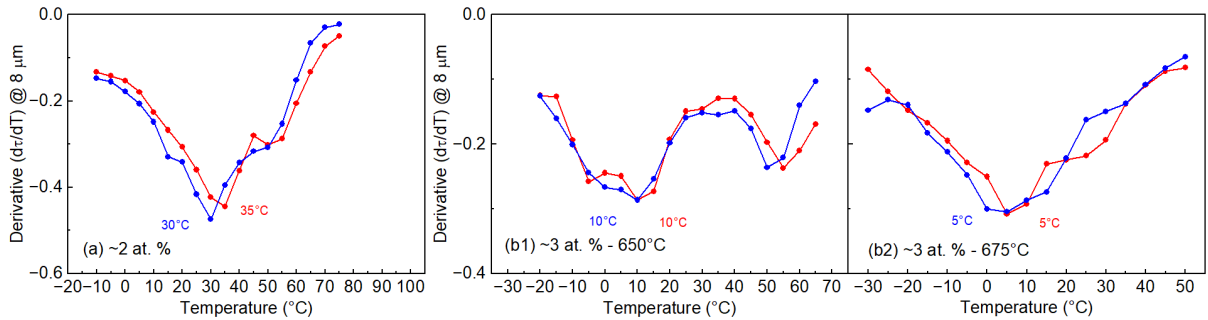

**Figure 7.** Thermal annealing and vacuum reduction studies in terms of transmittance derivative at 8 μm wavelength upon heating and cooling for (a) ~2 at.% WVO$_2$ sample annealed at 650°C and (b) ~3 at.% WVO$_2$ annealed at 650°C and 675°C followed by 3-hour vacuum reduction at 450°C.

3.3 IMT behaviors of optimally grown WVO$_2$ thin films

After the 50-nm W$_x$V$_{1-x}$O$_2$ thin films of three different doping levels are optimally grown with the aforementioned thermal oxidation, high-temperature annealing and surface reduction processes in low-O$_2$ furnace environment, XPS characterizations are carried out to determine the tungsten doping levels at 3 different locations of each film after 5 mins etching with 5 keV Ar$^+$ ions. Figure 8 shows the V2p and W4f elemental scans of all three WVO$_2$ samples, from which the tungsten doping levels are determined respectively to be 1.0, 2.1 and 2.7 at.% with standard deviation of ±0.1, ±0.2 and 0.2 at.% from three independent measurements.

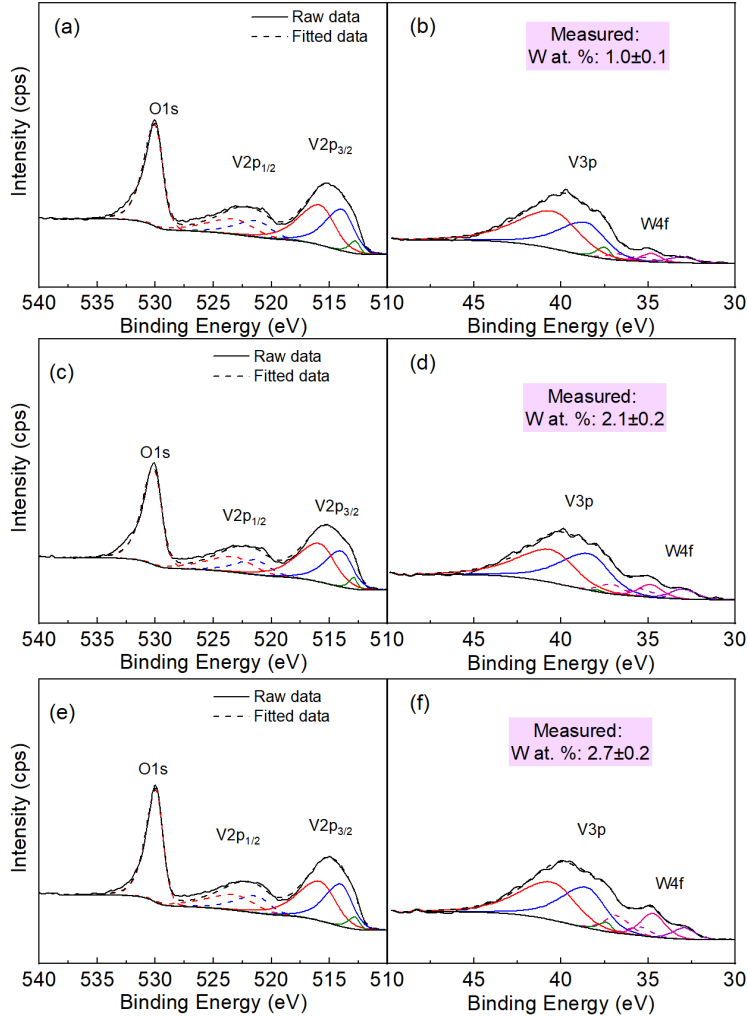

**Figure 8.** XPS characterization for determining tungsten doping levels of $W_xV_{1-x}O_2$ samples optimally grown in low-$O_2$ environment: (a) $x$ = 1.0 at.%, (b) 2.1 at.%, and (c) 2.7 at.%.

Figure 9 shows the temperature-dependent infrared transmittance spectra upon heating and cooling of all four 50-nm $W_xV_{1-x}O_2$ thin films of different doping levels ($x$ = 0, 1.0, 2.1, 2.7 at.%) optimally grown on UDSi wafers in low-$O_2$ furnace environment: (a) $x$ = 0 (undoped), (b) 1.0 at.%, (c) 2.1 at.%, and (d) 2.7 at.%. The spectra are captured with every 5°C interval except for the undoped $VO_2$ within IMT region where 2°C interval is used. Phase transition with significant variation in transmittance is clearly observed for all $WVO_2$ thin films. The infrared transmittance in the insulating phase (at lowest temperature measured) drops with more tungsten doping, while the transmittance in the metallic phase (at highest temperature measured) increases slightly with pronounced peaks around 10 μm wavelength in more heavily doped $WVO_2$ films due to higher temperatures required for sufficient annealing.

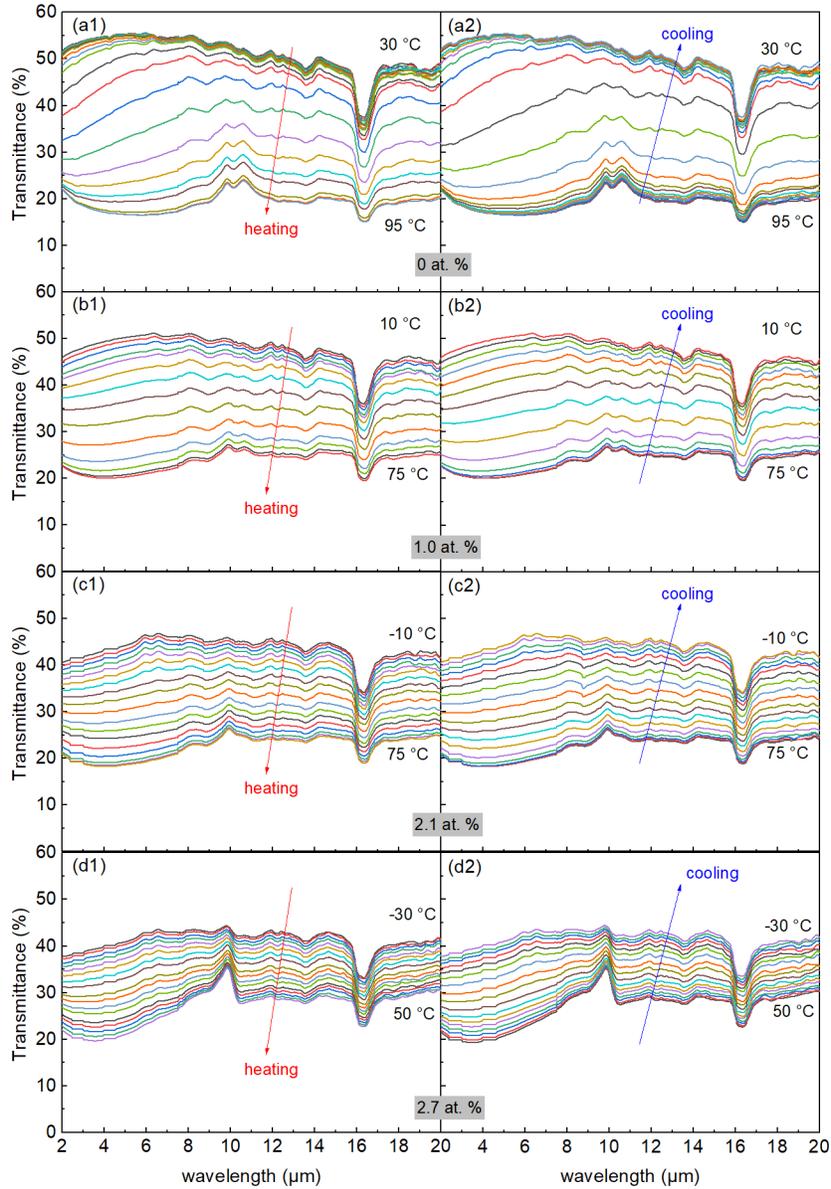

**Figure 9.** Temperature-dependent infrared transmittance spectra from −30°C to 95°C of all four 50-nm $W_xV_{1-x}O_2$ thin films of different doping levels optimally grown on UDSi wafers in low-$O_2$ furnace environment: (a) $x = 0$ (undoped), (b) 1.0 at.%, (c) 2.1 at.%, and (d) 2.7 at.%.

To better understand the IMT behaviors of all four $WVO_2$ films of different doping levels optimally grown in low-$O_2$ environment, spectral transmittance at 8 μm wavelength is plotted as a function of sample temperature upon both heating and cooling processes in Figure 10(a). Clearly, the insulator-to-metal phase transition shifts to lower temperatures with more tungsten doping. In addition, the thermal hysteresis between heating and cooling also becomes smaller from 10°C hysteresis with undoped $VO_2$ to almost none with 2.7 at.% tungsten doping. Figure 10(b) presents the derivative of the infrared transmittance at 8 μm wavelength with respect to sample temperature respectively upon heating and cooling processes for all four WVO$_2$ thin films. The IMT derivative valley moves to lower temperatures along with smaller magnitudes with higher tungsten doping. In particular, the IMT midpoint temperatures, at which there exists the largest derivative magnitude, are 72°C, 50°C, 35°C, and 5°C upon heating and 64°C, 50°C, 30°C, and 5°C upon cooling respectively for $W_xV_{1-x}O_2$ films with doping levels of x = 0, 1.0, 2.1, and 2.7 at.%. Temperature-dependent electrical resistivity is also measured for these four WVO$_2$ thin films optimally grown in low-O$_2$ furnace as shown in Figure 10(c), where IMT moving to lower temperature ranges with more tungsten doping is also observed. Notably, the $W_{0.027}V_{0.973}O_2$ exhibits only 4-fold change in resistivity due to its high tungsten doping, while the films with lower doping achieve about 40 times variation in resistivity upon phase transition. Figure 10(d) presents the resistivity derivative with respect to the sample temperature for all four WVO$_2$ thin films, where similar behaviors with consistent IMT midpoints upon heating and cooling as infrared transmittance are observed. Finally, a linear fitting is done on the IMT midpoint temperatures and is presented in Figure 10 (e) and (f). It shows the IMT temperature decreases by 23°C upon heating or 21°C upon cooling per at.% of tungsten doping, which agree well with literature reported values [44–47].

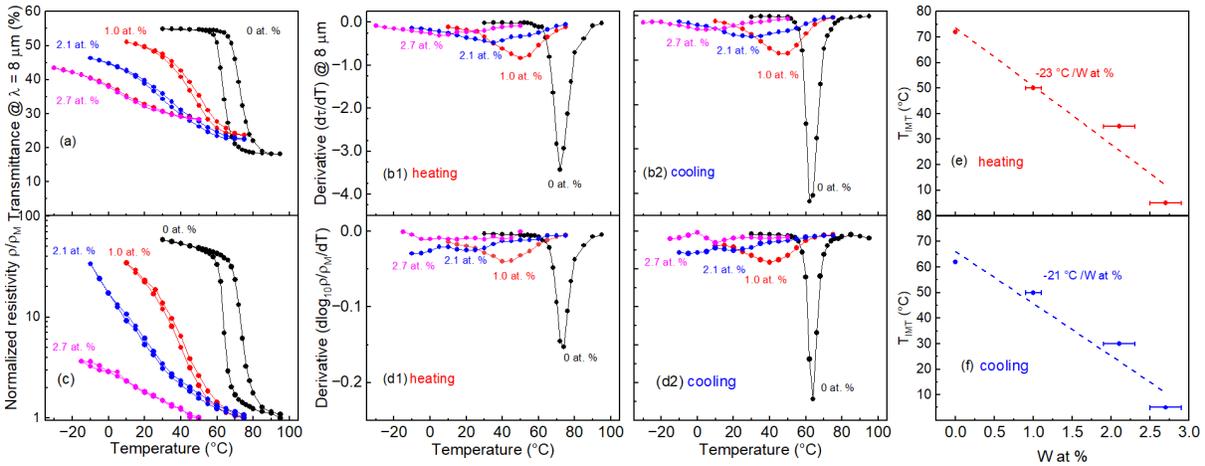

**Figure 10.** Insulator-to-metal transition behaviors of all four 50-nm $W_xV_{1-x}O_2$ thin films of different doping levels (x = 0, 1.0, 2.1 and 2.7 at.%) optimally grown on UDSi wafers via streamlined furnace oxidation, high-temperature annealing and surface reduction in low-O$_2$ environment: (a) temperature-dependent infrared transmittance at 8 μm wavelength and (b) its derivatives with respect to temperature upon heating and cooling; (c) temperature-dependent electrical resistivity normalized to its lowest value at metallic phase; (d) its derivatives; and transition temperature midpoint upon (e) heating and (f) cooling.

# 4. Conclusions

In summary, this work successfully demonstrated fabrication of high-quality WVO$_2$ thin films by oxidization of sputtered tungsten-vanadium alloyed thin films, high-temperature annealing, and vacuum reduction of surface over-oxides in low-O$_2$ furnace environment. The IMT behavior of undoped VO$_2$ is significantly improved compared to that oxidized in O$_2$-rich environment due to effective prevention of over-oxides with extremely low O$_2$ concentration confirmed by XRD. While surface over-oxide could be still formed at higher temperatures required for fully annealing WVO$_2$ thin films, vacuum reduction is implemented to improve the quality of WVO$_2$. Optimal oxidation, annealing and reduction conditions are determined, and the optimized WVO$_2$ thin films show lowered IMT temperature at the rate of –23°C/ at.% during heating and –21°C/ at.% during cooling, respectively. This cost-effective and scalable fabrication approach for high-quality WVO$_2$ thin films could lead to wide thermal and energy applications.

# CRedit authorship contribution statement

**Vishwa Krishna Rajan:** Conceptualization, Data curation, Formal analysis, Investigation, Methodology, Writing-original draft, Writing-review and editing. **Ken Araki:** Data curation, Writing-review and editing. **Robert Y. Wang:** Data curation, Writing-review and editing. **Liping Wang:** Conceptualization, Formal analysis, Funding acquisition, Investigation, Methodology, Supervision, Writing-review and editing.

# Acknowledgements

The work was supported by the National Science Foundation (NSF) under Grant No. CBET- 2212342. We are grateful to ASU Nanofab and Goldwater Center for using the fabrication and characterization facilities. V.K.R would like to thank ASU Graduate College and School for Engineering of Matter, Transport & Energy department for providing the financial support.

# Declaration of competing interest

The authors declare that they have no known competing financial interests of personal relationships that could have appeared to influence the work reported in this paper.